\documentclass{article}

\usepackage{microtype}
\usepackage{graphicx}
\usepackage{subfigure}
\usepackage{booktabs} 
\usepackage[linesnumbered,ruled]{algorithm2e}

\usepackage{amsmath,amsfonts,bm}

\def\eqref#1{equation~\ref{#1}}

\def\1{\bm{1}}

\def\vx{{\bm{x}}}

\DeclareMathAlphabet{\mathsfit}{\encodingdefault}{\sfdefault}{m}{sl}
\SetMathAlphabet{\mathsfit}{bold}{\encodingdefault}{\sfdefault}{bx}{n}

\usepackage{hyperref}

\usepackage[accepted]{icml2023}

\usepackage{amsmath}
\usepackage{amssymb}
\usepackage{mathtools}
\usepackage{hyperref}
\usepackage{amsthm}

\usepackage[capitalize,noabbrev]{cleveref}

\theoremstyle{plain}

\theoremstyle{definition}

\theoremstyle{remark}

\usepackage[textsize=tiny]{todonotes}

\icmltitlerunning{Meta-Learning the Inductive Bias}

\begin{document}

\twocolumn[
\icmltitle{Meta-Learning the Inductive Bias of Simple Neural Circuits}



\icmlsetsymbol{equal}{*}

\begin{icmlauthorlist}
\icmlauthor{William Dorrell}{yyy}
\icmlauthor{Maria Yuffa}{yyy}
\icmlauthor{Peter Latham}{yyy}
\end{icmlauthorlist}

\icmlaffiliation{yyy}{Gatsby Computational Neuroscience Unit, UCL}

\icmlcorrespondingauthor{William Dorrell}{dorrellwec@gmail.com}

\icmlkeywords{Neural Networks, Inductive Bias, Interpretability, Neuroscience, Spiking Neural Networks}

\vskip 0.3in
]



\printAffiliationsAndNotice{}  

\begin{abstract}

Training data is always finite, making it unclear how to generalise to unseen situations. But, animals do generalise, wielding Occam's razor to select a parsimonious explanation of their observations. How they do this is called their inductive bias, and it is implicitly built into the operation of animals' neural circuits. This relationship between an observed circuit and its inductive bias is a useful explanatory window for neuroscience, allowing design choices to be understood normatively. However, it is generally very difficult to map circuit structure to inductive bias. Here, we present a neural network tool to bridge this gap. The tool meta-learns the inductive bias by learning functions that a neural circuit finds easy to generalise, since easy-to-generalise functions are exactly those the circuit chooses to explain incomplete data. In systems with analytically known inductive bias, i.e. linear and kernel regression, our tool recovers it. Generally, we show it can flexibly extract inductive biases from supervised learners, including spiking neural networks, and show how it could be applied to real animals. Finally, we use our tool to interpret recent connectomic data illustrating its intended use: understanding the role of circuit features through the resulting inductive bias.

\end{abstract}

\section{Introduction}

Generalising to unseen data is a fundamental problem for animals and machines: you receive a set of noisy training data, say an assignment of valence to the activity of a sensory neuron, and must fill the gaps to predict valence from activity, Fig.~1A. This is hard since, without prior assumptions, it is completely underconstrained. Many explanations or hypotheses perfectly fit any dataset \citep{hume1748enquiry}, but different choices lead to wildly different outcomes. Further, the training data is likely noisy; how you sift signal from noise can heavily influence generalisation, Fig.~1B. 

Generalising requires prior assumptions about likely explanations of the data. For example, prior belief that small changes in activity lead to correspondingly small changes in valence would bias you towards smoother explanations, breaking the tie between options 1 and 2 in Fig.~1A. It is a learner's inductive bias that chooses certain, otherwise similarly well-fitting, explanations over others. Beyond this, the inductive bias is not rigorously defined. For the purposes of this paper we operationalise the inductive bias via the generalisation error: learners are inductively biased towards functions they generalise with low error.

\begin{figure}[h]
\begin{center}
\includegraphics[width=0.48\textwidth]{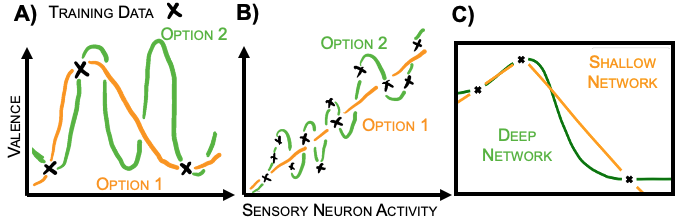}
\caption{\textbf{Generalisation Requires Prior Assumptions. A:} The same dataset is perfectly fit by many functions. \textbf{B:} Different assumptions about signal quality lead to different fittings. \textbf{C:} Training a 2 (shallow) or 8 (deep) layer ReLU network on the same dataset leads to different generalisations.}
\label{Fig: 1}
\end{center}
\vspace{-0.5em}
\end{figure}

The inductive bias of a learning algorithm, such as a neural network, can be a powerful route to understanding in both Machine Learning and Neuroscience. Classically, the success of convolutional neural networks can be attributed to their explicit inductive bias towards translation-invariant classifications \citep{lecun1998gradient}, and these ideas have since been successfully extended to networks with a range of structural biases \citep{bronstein2021geometric}. Further, many network features have been linked to implicit regularisation of the network, such as the stochasticity of SGD~\citep{mandt2017stochastic}, parameter initialisation~\citep{glorot2010understanding}, early stopping ~\citep{hardt2016train}, or low rank biases of gradient descent~\cite{gunasekar2017implicit}.

In neuroscience, the inductive bias has been used to assign normative roles to representational or structural choices via their effect on generalisation. For example, the non-linearity in neural network models of the cerebellum has been shown to have a strong effect on the network's ability to generalise functions with different frequency content \citep{xie2022task}. Experimentally, these network properties vary across the cerebellum, hence this work suggests that each part of the cerebellum may be tuned to tasks with particular smoothness properties. This is exemplary of a spate of recent papers applying similar techniques to visual representations \citep{bordelon2020spectrum, pandey2021structured}, mechanosensory representations \citep{pandey2021structured}, and olfaction \citep{harris2019additive}.

Despite the potential of using inductive bias to understand neural circuits, the approach is limited, since mapping from learning algorithms to their inductive bias is highly non-trivial. Numerous circuit features (learning rules, architecture, non-linearities, etc.) influence generalisation. For example, training two simple ReLU networks of different depth to classify three data points leads to different generalisations for non-obvious reasons, Fig.~1C. In a few cases analytic bridges have been constructed that map learning algorithms to their inductive bias. In particular, the study of kernel regression, an algorithm that maps data points to a feature space in which linear regression to labels is performed \citep{sollich1998learning, bordelon2020spectrum, simon2021neural}, has been influential: all the cited examples of understanding in neuroscience via inductive bias have used this bridge. However, it severely limits the approach: most biological circuits cannot be well approximated as performing a fixed feature map then linearly regressing to labels!

Here, we develop a flexible approach that is able to learn the inductive bias of essentially any supervised learning algorithm. It follows a meta-learning framework: an outer neural network (the meta-learner) assigns labels to a dataset, this labelled dataset is then used in the inner optimisation to train the inner neural network (the learner). The meta-learner is then trained on a meta-loss which measures the generalisation error of the learner to unseen data. Through gradient descent on the meta-loss, the meta-learner meta-learns to label data in a way that the learner finds easy to generalise. These easy-to-generalise functions form a description of the inductive bias, since easy-to-generalise functions are those that learners generalise appropriately from few training points. They are therefore functions the network would regularly use to explain finite datasets. In sum, networks are inductively biased towards easy-to-generalise functions.

To our knowledge, the most related work is \citet{li2021meta}. Li et al. view sets of neural networks, trained or untrained, as a distribution over the mapping from input to labels. They fit this distribution by meta-learning the parameters of a gaussian process which assigns a label distribution to each input. This provides an interpretable summary of fixed sets of network. In our work we do something very different: rather than focusing on a fixed, static set of networks, we find the inductive biases of learning algorithms via meta-learning easily learnt functions. We recommend potential users to consider both approaches.

In the following sections we describe our scheme, and validate it by comparing to the known inductive biases of linear and kernel regression. We then extend it in several ways. First, networks are inductively biased towards areas of function space, not single functions. Therefore we learn a set of orthogonal functions that a learner finds easy to generalise, providing a richer characterisation of the inductive bias. Second, we introduce a framework that asks how a given design choice (architecture, learning rule, non-linearity) effects the inductive bias. To do that, we assemble two networks that differ only by the design choice in question, then we meta-learn a function that one network finds much easier to generalise than the other. This can be used to explain why a particular circuit feature is present. We again validate both schemes against linear and kernel regression. Finally we show our tool's flexibility in a series of more adventurous examples: we validate it on a challenging differentiable learner (a spiking neural network); we show it works in high-dimensions by meta-learning MNIST labels; we highlight its explanatory power for neuroscience by using it to normatively explain patterns in recent connectomic data; and we extract inductive biases using  gradient-free techniques, demonstrating a method that could be used to extract animals' inductive biases.

\begin{figure}[!h]
\begin{center}
\includegraphics[width=0.39\textwidth]{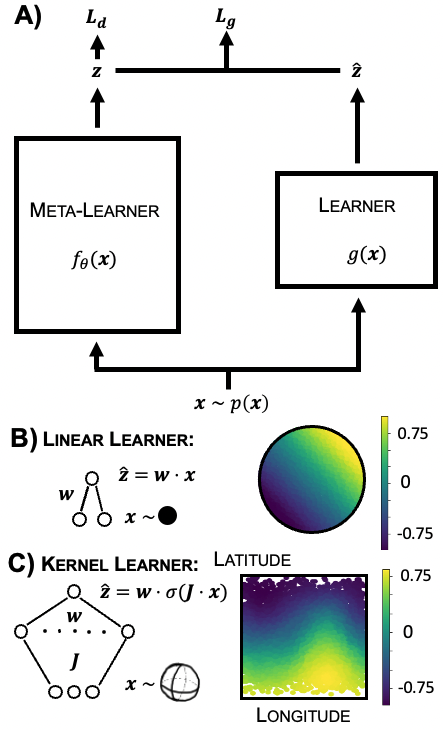}
\caption{\textbf{Meta-Learning the Inductive Bias. A:} The meta-learner labels a dataset, part of which is used to train the learner. Gradient descent is performed on a loss made of the learner's generalisation error on unseen data ($\mathcal{L}_g$), and the Sinkhorn divergence between the meta-learner's label distribution and a target distribution ($\mathcal{L}_d$), here choosen to be uniform from $-1$ to $1$. \textbf{B:} The meta-learner learns a linearly separable labelling of data sampled from a circle for a ridge regression learner. The plot (and similar future plots) shows the datapoints - the circle - coloured by the assigned meta-learner label. \textbf{C:} For a kernel regression learner and data sampled from the surface of a sphere, the meta-learner's labelling is very close to the predicted spherical harmonic ($99\%$ of norm within first order harmonics).}%
\label{Fig: 2}
\end{center}
\vspace{-1.5em}
\end{figure}

\section{Meta-Learning the Inductive Biases}

Our main contribution is a meta-learning framework for extracting the inductive bias of differentiable learning algorithms, Fig.~2A, that we describe in this section. In the outer-loop a neural network, the meta-learner, assigns labels to inputs sampled from some distribution, hence creating the real-world function that our circuit of interest will try to learn. The inner-loop learning algorithm, the learner, is the circuit whose inductive bias we want to extract; for example, a biological sensory processing circuit that assigns valences to inputs. When provided with a training dataset of inputs and labels the learner adjusts its parameters according to its internal learning rules. Then the generalisation error of the trained learner is measured on a held-out test set, and this is used as the meta-loss to train the meta-learner. This process repeats, reinitialising then retraining the learner at every iteration, iteratively developing the meta-learner's weights, until the meta-learner is labelling the data in a way that the learner finds easy to generalise after training on a few datapoints (we used $\sim$30). Thus, the meta-learner has extracted a function towards which the learner is inductively biased.

As just outlined, the meta-learner will find the easiest-to-generalise function, usually the one that assigns all inputs the same label. To avoid this trivial function, we introduce a term in the meta-loss that forces the distribution of labels to take a particular (non-constant) form. Specifically, it measures the Sinkhorn divergence between the meta-learner's label distribution and a uniform distribution from $-1$ to $1$ (other divergences also work, Appendix B, or other methods like forcing the label distribution's variance to be 1). The full pseudocode is in Algorithm~\ref{Psuedocode}.

Our meta-learner must learn a function that the learner can generalise. To enable the meta-learner to learn all functions the learner might plausibly generalise well, its function class could usefully be a superset of the learner's. Therefore, we choose the meta-learner's architecture to be a slightly larger version of the learner's (though, beyond this, our findings appear robust, Appendix D). 

Just like any other deep learning approach, the meta-learner's hyperparameters (learning rate etc.) must be chosen so that it optimises well. On the other hand, the learner's hyperparameters are chosen as part of specifying the learner - changing the learning rate or initialisation correspond to changing its inductive bias, so the method does and should extract different functions as these parameters vary.

\begin{algorithm}
    \SetKwInOut{Input}{Input}
    \SetKwInOut{Output}{Output}
    
    Initialise meta-learner: $f_\theta(\vx)$

    \While{Step count $<$ Total}
    {
        Generate dataset from input distribution: $\vx \sim p(\vx)$ \\
        Label using metalearner: $z = f_\theta(\vx)$ \\
        Split inputs and labels into test \& train sets: $\mathcal{D}_{Tr}$ \& $\mathcal{D}_{Te}$ \\
        Train leaner using $\mathcal{D}_{Tr}$ giving trained learner: $g(\vx)$ \\
        Label $\mathcal{D}_{Te}$ using trained learner: $\hat{z} = g(\vx)$\\
        Compute the learner's generalisation error: $\mathcal{L}_g = \sum_i (z_i - \hat{z}_i)^2$ \\
        Compute the Sinkhorn divergence of metalearner's labels from uniform $[-1, 1]$: $\mathcal{L}_d$ \\
        Take $\theta$ gradient step on meta-loss: $\mathcal{L} = \mathcal{L}_g + \mathcal{L}_d$
    }
    \caption{Meta-Learning the Learner's Inductive Bias}
\label{Psuedocode}
\end{algorithm}

We validate our scheme by meta-learning sensible functions for linear and kernel learners, whose inductive biases are known. First, for ridge regression on data sampled from a 2D circle the meta-learner assigns linearly separable labels, Fig.~2B; exactly the labels linear circuits easily generalise.

Next, we meta-learn kernel ridge regression's inductive bias. Kernel regression involves projecting the input data through a fixed mapping to a feature space (e.g. the last hidden layer of a fixed neural network) and performing linear regression from feature space to labels. \citet{bordelon2020spectrum} show that the inductive bias of kernel regression can be understood through the kernel eigenfunctions ($\{v_i(\vx)\}$ with eigenvalue $\{\lambda_i\}$). These are defined on input distribution $p(\vx)$ via a kernel $k(\vx,\vx')$ that measures the similarity of two inputs in feature space:
\begin{equation}
\int k(\vx,\vx') v_i(\vx')dp(\vx') = \lambda_iv_i(\vx).
\end{equation}
The algorithm is inductively biased towards higher eigenvalue eigenfunctions; i.e., fewer training points are needed to reach a given generalisation error when fitting high vs.~low eigenvalue eigenfunctions. General functions can be understood by projecting onto the eigenbasis. Hence our meta-learner, in searching for kernel regression's easiest-to-generalise non-constant function, should choose the highest eigenvalue eigenfunction.

\begin{figure*}[!h]
\begin{center}
\includegraphics[width=0.8\textwidth]{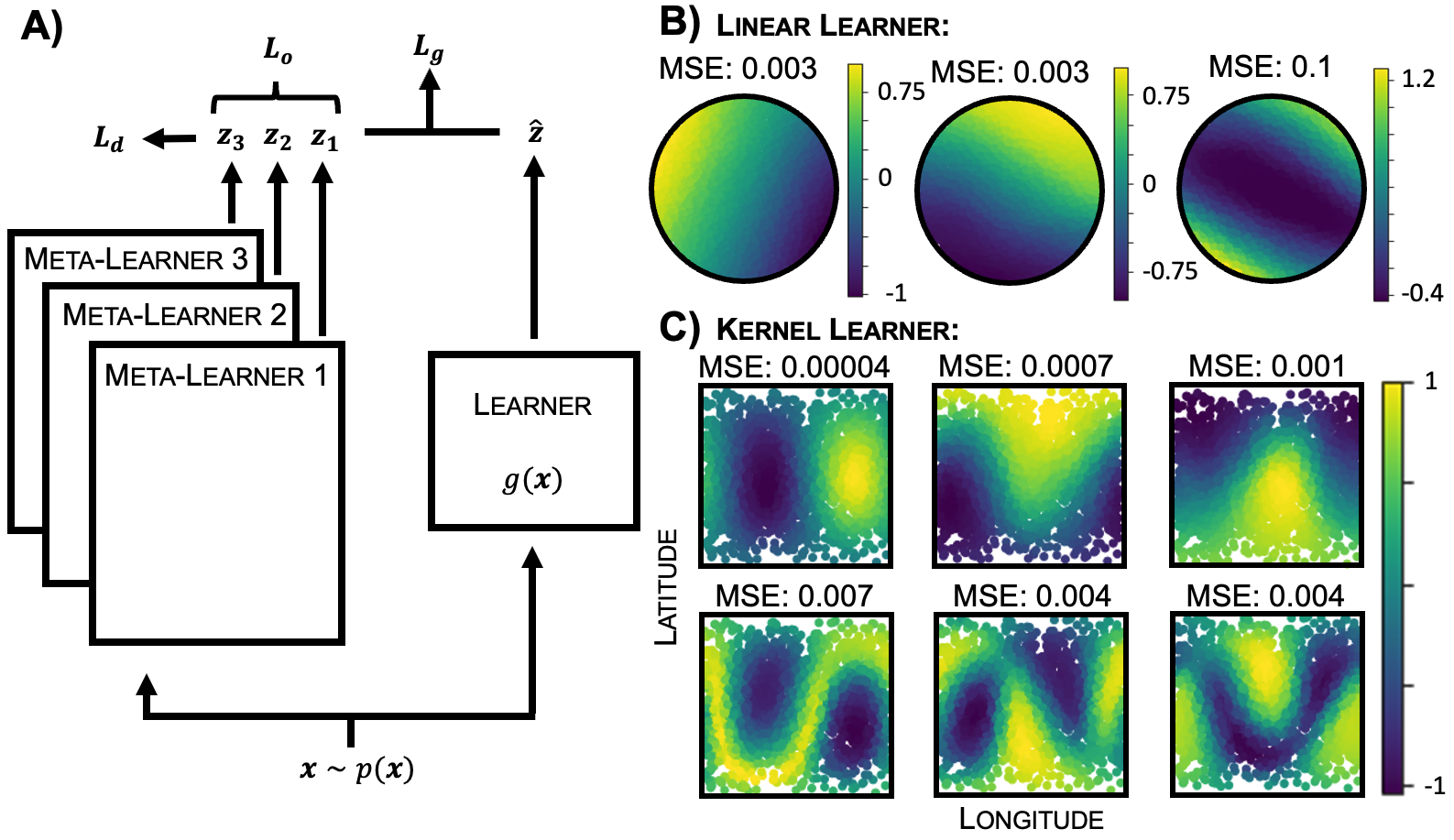}
\caption{\textbf{Meta-Learning Many Functions. A:} We learn many meta-learners, each of which has to label orthogonally to all previous meta-learners. \textbf{B:} For a linear learner the meta-learners learn two orthogonal linear functions and an orthogonal but hard to learn third function. \textbf{C:} For a kernel learner we learn 6 meta-learners, the first 3 approximate well first order spherical harmonics (96\% norm overlap), and the next 3 second order spherical harmonics (91\% norm overlap), as predicted.}
\label{Fig: 3}
\end{center}
\vspace{-0.5em}
\end{figure*}

To test this, we meta-learn the inductive bias of a two-layer neural network with fixed first layer weights. We sample data uniformly from the sphere and randomly connect a large hidden layer of ReLU neurons to the three input neurons. The elements of this random weight matrix are drawn \textit{iid} from a standard normal, and the learning algorithm performs ridge regression on the hidden layer activities. Previous work has analytically derived the kernel for this network, and computed its eigenfunctions \citep{cho2009kernel, mairal2018machine}, which are spherical harmonics\footnote{Spherical harmonics are an orthonormal basis of functions of increasing frequency defined on the sphere, analgously to sines and cosines on the circle.}. The higher the frequency of the spherical harmonic the lower its eigenvalue. Matching this, our network meta-learns one of the set of lowest frequency spherical harmonics, Fig.~2C.

\section{Meta-Learning Areas of Function Space}\label{sec:orthog}

Having validated our tool on some simple test cases, we now extend it to find a richer characterisation of the inductive bias. A given learning algorithm is inductively biased towards areas of function space, not just one particular function. To gain access to this larger space, we learn a series of meta-learners. The first of these is exactly as described above, 
then we iteratively introduce additional meta-learners. To ensure each meta-learner learns a new aspect of the inductive bias we add a term to the meta-loss that penalises the square of the dot product between the current meta-learner's labelling and that of all the previously trained meta-learners, Fig.~3A. On a dataset $\{\vx_n\}$:
\begin{equation}
    \mathcal{L}_{\text{Orthog}} = \sum_i \bigg(\sum_n f_{\theta_i}(\vx_n)f_{\theta'}(\vx_n)\bigg)^2
\end{equation}
for each previous meta-learners $f_{\theta_i}(\vx)$ and the current meta-learner $f_{\theta'}(\vx)$. From the learner's perspective nothing has changed, at each meta-step it simply learns to fit the meta-learner that is currently being trained. But each additional meta-learner must discover an easy-to-generalise function that is orthogonal to all previous meta-learners.

We again validate this scheme on linear and kernel regression. For linear regression of 2D data the meta-learners learn two orthogonal linear labellings, then a third orthogonal function that the learner struggles to generalise, as expected, Fig.~3B. For the kernel regression network we described previously theory predicts that the meta-learners should learn the eigenfunctions in decreasing order of their eigenvalue. We find this to a good approximation, Fig.~3C, learning approximations of, first, the three first order spherical harmonics, and second, three linear combinations of the second order harmonics.

\begin{figure*}[!h]
\begin{center}
\includegraphics[width=0.85\textwidth]{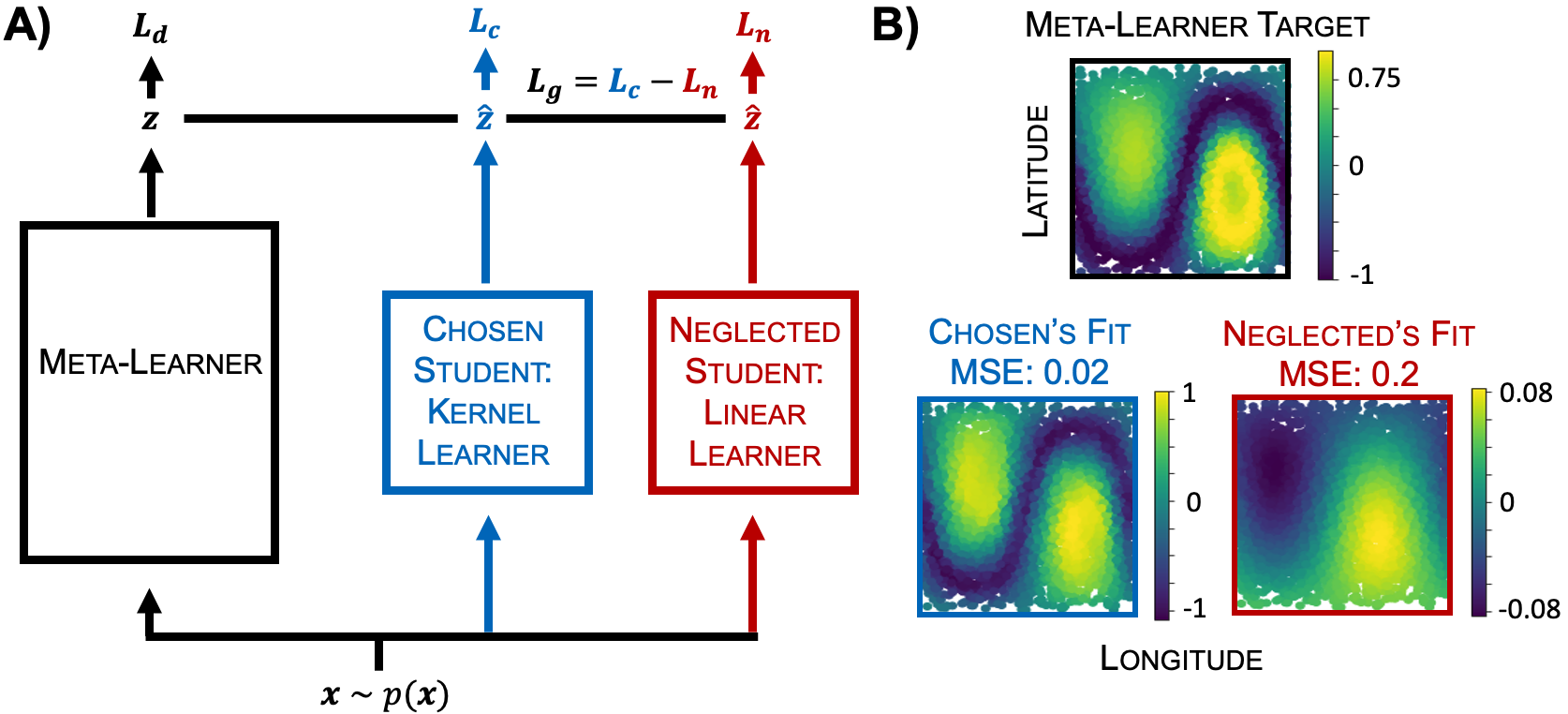}
\caption{\textbf{Meta-learning design choice impact. A:} Labellings are learnt so a chosen student generalises but a neglected one doesn't. \textbf{B:} The meta-learner finds a non-linear labelling on which kernel regression generalises an order of magnitude better than linear regression.}
\label{Fig: 4}
\end{center}
\end{figure*}

For linear classifiers (e.g. linear and kernel regression) the full set of orthogonal functions explains the entire inductive bias, since the average generalisation error on any target function is the sum of the errors on each kernel eigenfunctions weighted by projection of the target function onto that eigenfunction. This will not in general be true, since for a non-linear classifier the generalisation error on a function $f(\vx) = f_a(\vx) + f_b(\vx)$ does not equal the generalisation error on $f_a(\vx)$ plus that on $f_b(\vx)$. Nonetheless, we expect the set of orthogonal functions will be a helpful guide to a network's inductive bias, even for non-linear classifiers.

\section{Isolating a Design Choice's Impact}

Our work is motivated by the desire to understand how design choices in learning algorithms - such as architecture or non-linearities - lead to downstream generalisation effects, particularly in biological networks. One additional setting which we have found useful is to compare two networks with some difference between them, and learn functions that one of the networks finds much easier to generalise than the other. In this way, we can build intuition for the impact of design features on the inductive bias. 
To illustrate this we again create a meta-learner that labels data, but this time the labels are used to train two learners. We then train the meta-learner so that one learner (the chosen student) is much better at generalising than the other (neglected student). This is done by minimising the generalisation errors of the chosen student minus the neglected student, Fig.~4A.  Validating this approach on well understood algorithms, we show that it can find functions that a kernel regression algorithm is able to learn better than linear regression, Fig.~4B, i.e. a non-linearly separable function.

This illustrates some of the games that can be played in this setting. For example, you could play a co-operative game, in which you meta-learn a function that a set of learners all find easy to generalise, and each learner could have different connectivity matrices to match the distribution in real animals, ensuring the tool does not over-fit to some specific details. However as the losses become more complex training becomes harder, for example this adversarial setting between chosen and neglected student is hard to make robust if the two learners are relatively similar.

\section{Applying our Framework}

\begin{figure*}[h]
\begin{center}
\includegraphics[width=0.83\textwidth]{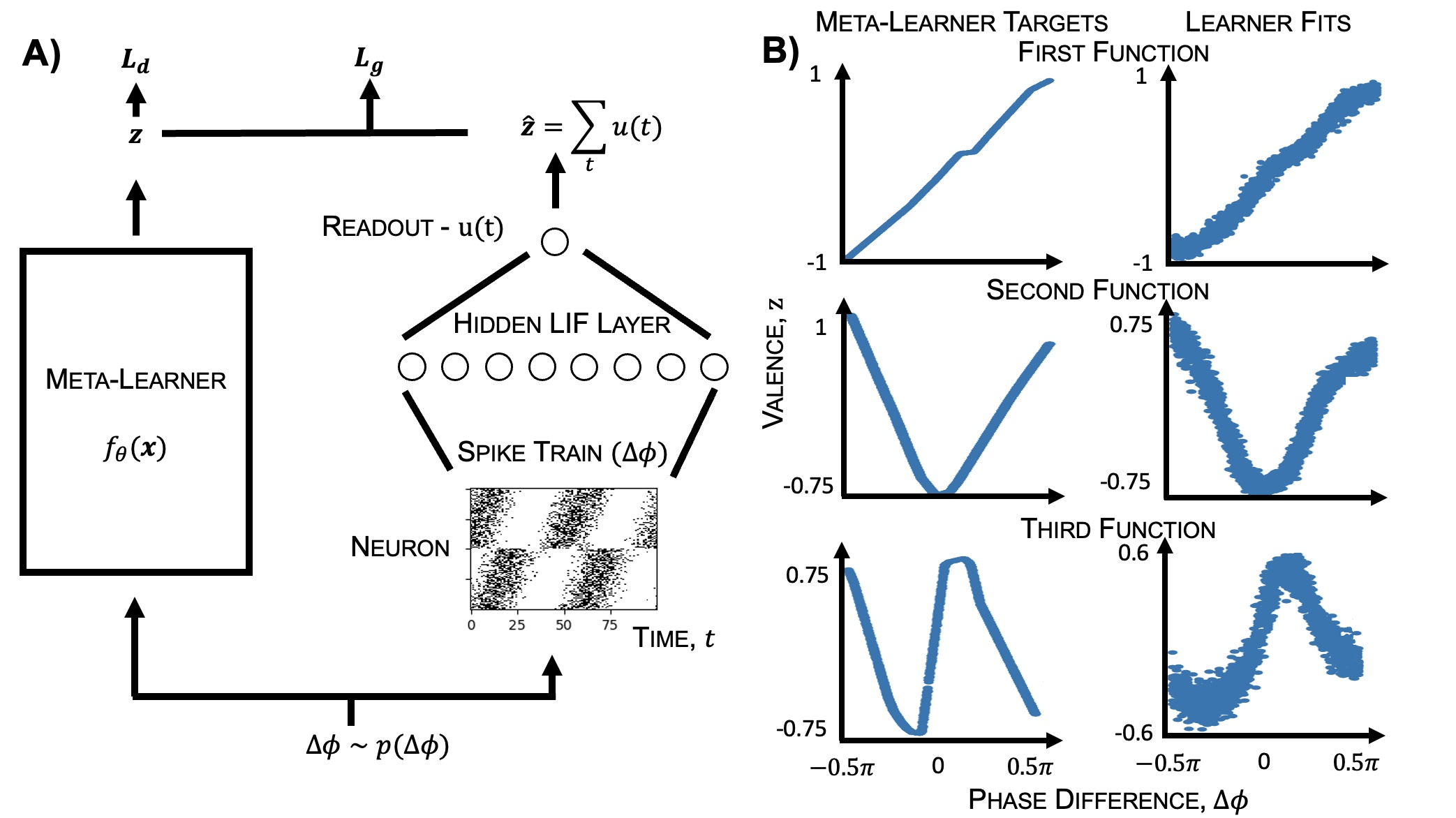}
\caption{\textbf{Meta-learning through a Spiking Network. A:} Labellings are learnt that the spiking network, with weights trained via surrogate gradient descent, finds easy to generalise. Phase differences, $\Delta \phi$, are sampled uniformly and used to generate spike trains by sampling from a poisson process with the following rates: for half the neurons $r_n = \frac{r_{\text{max}}}{2}(1 + \sin(t + \theta_n))^2$, where $n$ is a neuron index and $\theta_n$ are uniformly sampled offsets; for the other half we add a phase shift:  $r_n = \frac{r_{\text{max}}}{2}(1 + \sin(t + \theta_n + \Delta\phi))^2$. These populations represent sensory neurons in the two ears, and $\Delta\phi$ is the interaural phase difference. This activity feeds into a population of linear-integrate-and-fire neurons, then to a readout linear-integrate neuron. The valence assigned is the sum of the readout's activity over time. \textbf{B:} We learn three orthogonal meta-learners (as in section~\ref{sec:orthog}). The spiking network finds it easiest to learn low frequency functions. Left: the meta-learner's target function. Right: the spiking network's labelling. The spiking network captures the main behaviour, but increasingly poorly.}
\label{Fig:Spike}
\end{center}
\vspace{-1em}
\end{figure*}

So far we have developed and tested a suite of tools for extracting the inductive bias of learning algorithms. We now apply them to networks whose inductive bias cannot be understood analytically. Specifically: we show our method works on a challenging differentiable learner, a spiking neural network; we validate our method on a high-dimensional MNIST example; we illustrate how our tool can give normative explanations for biological circuit features, by meta-learning the impact of connectivity structures on the generalisation of a model of the fly mushroom body; and we demonstrate a method to extract animals' inductive biases.  Our tool is flexible: by taking gradients through the training procedure we can meta-learn inductive biases for networks trained using PyTorch, for example. We share all our code\footnote{Code at \url{https://github.com/WilburDoz/Meta\_Learning\_Inductive\_Bias}} which should be easily adapted to networks of interest.

\subsection{Spiking Neural Network}

The brain, unlike artificial neural networks, computes using spikes. `How?' is an open question. A recent exciting advance in this area is the surrogate gradient method, which permits gradient based training of spiking neural networks by smoothing the discontinuous gradient~\citep{neftci2019surrogate, zenke2021remarkable}. We make use of this development to meta-learn the inductive bias of a spiking network, providing a challenging test case for our method.

We study a modification of a model developed for a recent tutorial \citep{goodman2022cosyne, zenke2019spytorch}, which is trained to assign a label to an incoming spike train. The network is a model of an interaural phase difference detection circuit. The input spike train is parameterised by a phase difference, $\Delta\phi$, that generates two sets of spike trains, one in each ear, Fig.~\ref{Fig:Spike}A. These spikes are processed through a hidden layer of linear-integrate-and-fire neurons (LIF), before reaching a classification layer. A real-valued valence is assigned by summing the output neuron's activity over the trial. The meta-learning framework is as before: the meta-learner assigns valences to input phase differences, these labels are used to train the spiking network by surrogate gradient descent, then the meta-learner is trained to minimise the learner's generalisation error and a distribution loss. Our method works well, finding a simple smoothness prior, Fig.~\ref{Fig:Spike}B.

\begin{figure}[h]
\begin{center}
\includegraphics[width=0.37\textwidth]{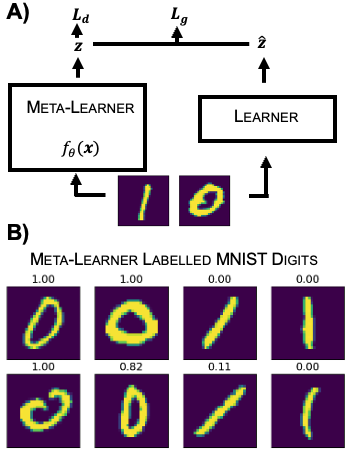}
\caption{\textbf{Meta-Learning on MNIST A:} A meta-learner receives MNIST 0s and 1s, and assigns labels, bounded between 0 and 1, that have high variance and can be easily generalised by the learner. \textbf{B:} 99\% of digits are assigned a label, shown in title, consistent with MNIST class.}
\label{fig:MNIST}
\end{center}
\vspace{-1.5em}
\end{figure}

\subsection{A High-Dimensional MNIST Example}

Next, we test out method on a high-dimensional input dataset. Thus far, to visualise our results, we have only considered low dimensional input data. We demonstrate that our method continues to work in high-dimensions by applying it to a dataset made of the 0 and 1 MNIST digits  \citep{lecun1998mnist}. We meta-learn a labelling of this dataset that a simple convolutional neural network finds easy to generalise. Our meta-learner's architecture is also a convolutional neural network whose outputs are bounded between 0 and 1, and the meta-learner must learn an easy-to-generalise labelling with high variance. We find that the meta-learner consistently rediscovers the MNIST digits within the dataset, separating each digit into its own class, figure~\ref{fig:MNIST}. 

While successful, this example highlights the major challenge of applying our method in high-dimensions. To properly probe the inductive bias of a CNN would require finding an easily-generalised set of functions from images to labels. How would you intepret such functions? We return to this concern in the discussion; fortunately, we show in the next section that our method can be practically useful when combined with meaningful projections of the data.

\subsection{Assigning Normative Roles to Connectomic Data}

A large maturing source of neuroscience data is connectomics (a list of which neurons connect to one another). However, there is currently a dearth of methods for interpreting this data \citep{litwin2019constraining}. In this section, we show our tool can be used to give normative roles to connectomic patterns through their induced inductive bias. We study a model of the fly mushroom body, a beautiful circuit that fruit flies use to assign valence to odours \citep{aso2014neuronal, hige2018can}, for which connectomic data has recently become available \citep{zheng2018complete,zheng2022structured}.

\begin{figure*}[h]
\begin{center}
\includegraphics[width=0.9\textwidth]{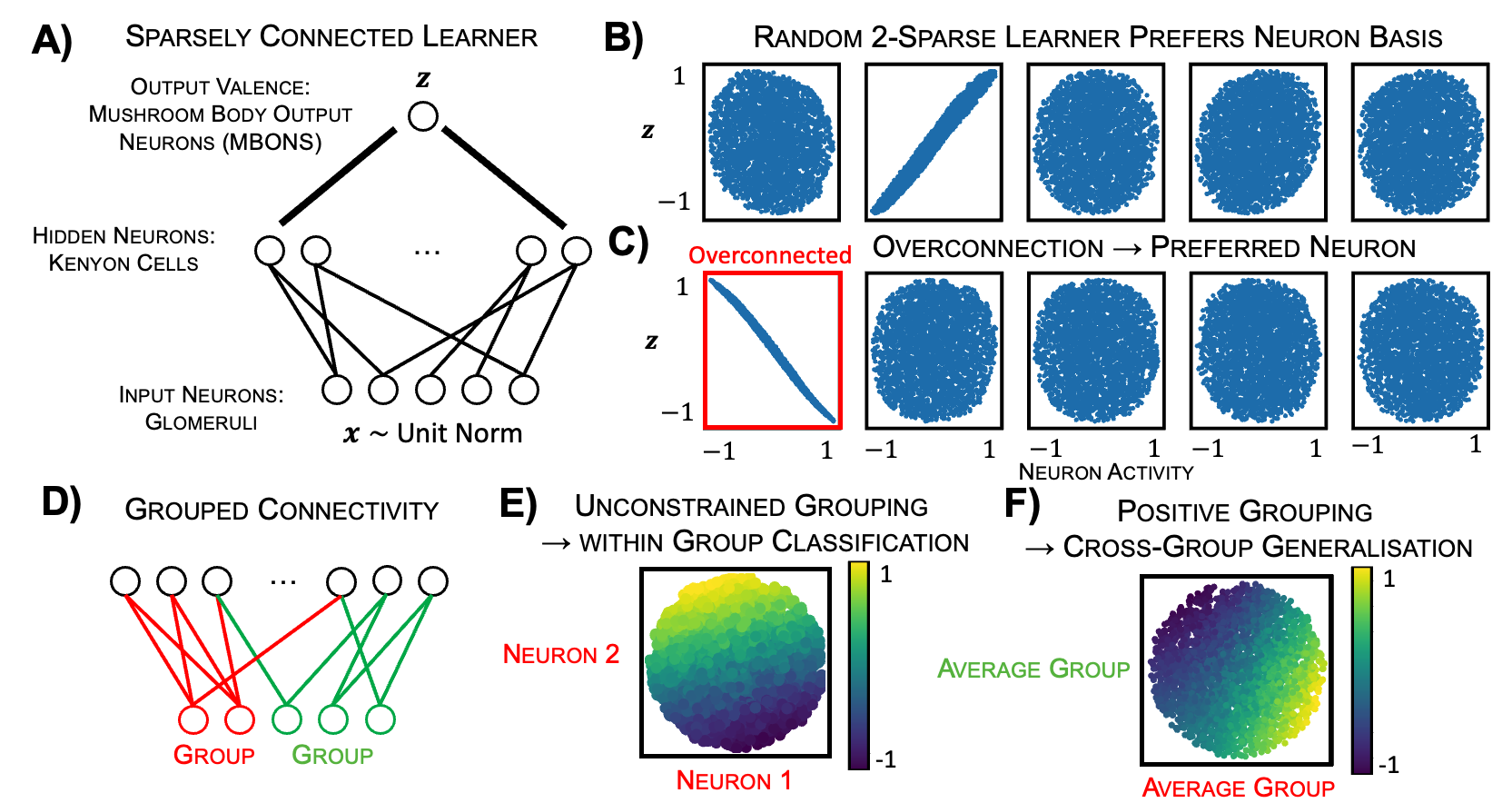}
\caption{\textbf{Understanding Connectivity via Inductive Bias. A:} We model the fly mushroom body as a ReLU network with five inputs and one large hidden layer. Each hidden neuron is connected to two of the five input neurons, and we study three networks in which these two connections are chosen differently, either randomly (\textbf{B}), biased (\textbf{C}) or grouped (\textbf{D} - \textbf{F}). \textbf{B:} Shows the easiest-to-generalise function for a network in which each hidden neuron is connected to two random inputs. Each of the five plots show this function projected against one input neuron's activity. As can be seen, the labelling depends on only one neuron's activity, second from left. \textbf{C:} In the overconnected setting each hidden neuron still connects to two inputs, but there is a strong bias towards connecting to the first, highlighted neuron. As a result, the meta-learner settles on a labelling that depends only on this neuron's activity. \textbf{D:} We explore the impacts of group connectivity, in which the input neurons are divided into two groups, and hidden neurons tend to be connected to two neurons from the same group. \textbf{E:} We train the meta-learner, and find that it's labelling depends only on neurons within the same group. The plot shows the projection of the datapoints into a subspace defined by the two input neurons in the red group. The labelling depends linearly on position within this subspace. \textbf{F:} However, if the input-hidden connections are constrained to be positive, the meta-learner's labelling depends only on the average activity within each group, i.e. if one member of a group increases the output, so do all members; hence, the function generalises across group members. The plot shows the two dimensional subspace defined by the mean activity of each of the groups of inputs.}
\label{Fig:Fly}
\end{center}
\end{figure*}

Odorants trigger a subset of the fly's olfactory receptors. These activations are represented in a small glomerular population (input neurons), projected to a large layer of Kenyon cells (hidden neurons), then onwards to output neurons that signal various dimensions of the odour's valence, Fig.~\ref{Fig:Fly}A. An error signal is provided if the fly misclassifies a good odour as bad, or vice versa, allowing the fly to update its weights and learn appropriate responses. Classically, the input-to-hidden connectivity was assumed random; i.e., each hidden neuron connects to a few randomly selected inputs. However, connectomic data has shown two patterns; first, hidden neurons preferentially connect to some inputs, and second, there are input groupings - if a hidden neuron connects to one member of a group it likely connects to many, Fig~\ref{Fig:Fly}D \citep{zheng2018complete, zheng2022structured}. \citet{zavitz2021connectivity} tested networks with this connectivity on a battery of tasks and found that, compared to random, (1) they were better at identifying odours that activated over-connected inputs, and (2) they generalised assigned valence across a group (i.e. if you assign high valence to the activation of one neuron, you do the same for other neurons in the same group).

We used our tool to verify and develop these findings by examining the effect of different connectivity patterns on the inductive bias of a sparsely-connected model of the mushroom body, Fig.~\ref{Fig:Fly}A, Appendix E. As a baseline, fully connected networks are biased towards smooth functions, appendix~\ref{app:ReLU}, the simplest being those that assign valence based on one direction in the input space: high at one end, low at the other, like in Fig.~\ref{Fig: 2}B - C. However, which direction is unimportant; they're all equally easy to learn. Sparsity breaks this degeneracy, aligning the easiest to learn functions with the input neuron basis, figure~\ref{Fig:Fly}B. As such, sparse connectivity, which is ubiquitous in neuroscience, ensures the fly is best at assigning labels based on the activity of small collections of neurons. 
Next, we introduced the observed connectomic structure. Biasing the connectivity, so some inputs have more connections than others, broke the degeneracy amongst neuron axes. The networks were, as expected, best at generalising functions that depended on the activity of overconnected inputs, figure~\ref{Fig:Fly}C, matching \citet{zavitz2021connectivity}. Finally we introduce connectivity groups, figure~\ref{Fig:Fly}D. Without additional changes this does little, the neuron basis is still preferred and, unlike~\citet{zavitz2021connectivity}, generalisation across inputs is not observed, figure~\ref{Fig:Fly}E. Only when we additionally constrain the input-to-hidden connections to be excitatory (i.e. positive) do we see that the circuit becomes inductively biased towards functions that generalise across groups of inputs, figure~\ref{Fig:Fly}F. In retrospect this can be understood intuitively: positive weights and grouped connectivity ensure that a hidden neuron that is activated by one input will also be activated by other group members, encouraging generalisation. This effect is removed by permitting negative weights, which let members of the same group excite or inhibit the same hidden neuron.

Thus, we verify and extend the findings of~\citet{zavitz2021connectivity}. We, first, show the results hold without needing to hand-design a battery of tasks. This avoids a potential flaw in the approach of~\citet{zavitz2021connectivity}: you may reach an incorrect conclusion simply because your battery is not comprehensive enough! (though in this case we agree wholeheartedly with the conclusions of~\citet{zavitz2021connectivity}) Our method avoids this problem by meta-learning the appropriate tasks. Second, we study networks in which both sets of weights are trained, rather than just readout, and, third, we emphasise the role of sparsity in aligning easy-to-learn functions with the neuron basis, something not studied in Zavitz et al. In sum, this example illustrates how our tool can be used to normatively understand elements of circuit design.

\subsection{Animals' Biases via Black-box Optimisation}

\begin{figure}[h]
\begin{center}
\includegraphics[width=0.4\textwidth]{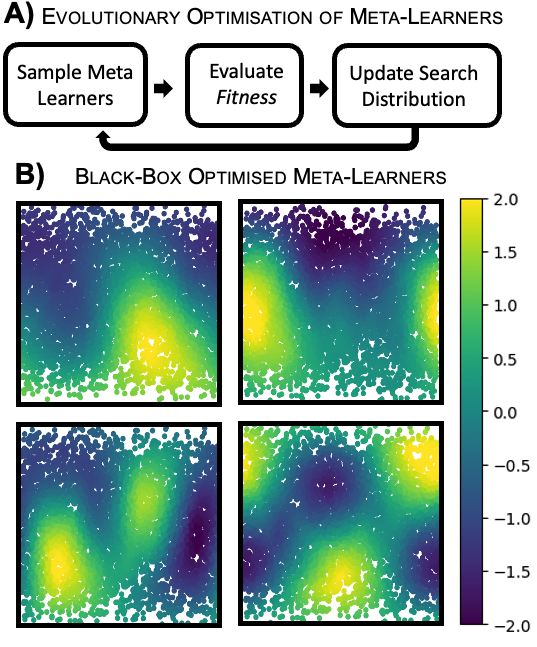}
\caption{\textbf{Black-Box Optimisation of Meta-Learners. A:} Evolutionary optimisation of meta-learners involves sampling a set of meta-learner parameters, evaluating their fitness, and using these evaluations to update the search distribution (using a scheme described in \citet{lange2022discovering}). The fitness is the full loss, including the learner's generalisation error, the orthogonality loss (eqn. 2), and a label distribution loss. \textbf{B:} Trained on the kernel regression problem from section 2 it recovers functions of increasing frequency, though misses one of the first order spherical harmonics.}
\label{Fig: 8A}
\end{center}
\vspace{-1em}
\end{figure}
In our final, slightly kooky, experiment we modify and test an approach to directly interface with an animal. After all, we are interested in animals' inductive biases and how these arise from neural circuitry; in many settings a better way to probe these is to avoid modelling the circuit entirely, and to simply extract 
easy-to-generalise functions from the animal's behaviour. This can be done by replacing the inner learner with a real animal trained on a labelled dataset from the meta-learner, then tested on new datapoints. This quickly runs into a problem: we cannot take gradients through the computations of a living animal. However, we can avoid needing these gradients by using evolutionary optimisation to train the meta-learner, which requires only evaluations of the generalisation error. 

In figure 8 we demonstrate the efficacy of this scheme. We use the Discovered Evolution Strategy from \citet{lange2022discovering} (implemented in evosax \citep{lange2022evosax}), to learn functions that the simple kernel ridge regression circuit discussed in section 2 generalises well. We sample a distribution of meta-learners, measure the learners generalisation error on these, and update the meta-learner distribution accordingly. We are eager to try this approach experimentally.

\section{Discussion \& Conclusions}

We presented a meta-learning approach to extract the inductive bias of differentiable supervised learning algorithms, which we hope will be useful in normatively interpreting the role of features of biological networks. This approach required few assumptions beyond those that make the inductive bias an interesting way to conceptualise a circuit in the first place. We required, first, that the input data distribution was specified, and, second, the circuit must be interpretable as performing supervised learning. We will discuss these requirements and ways they could be relaxed; regardless, it is heartening that any circuit satisfying these will, in principle, suffice. The analytic bridge between kernel regression and its inductive bias \citep{bordelon2020spectrum, simon2021neural} has already found multiple uses in biology in just a few years \citep{bordelon2021population, pandey2021structured, harris2019additive, xie2022task}, despite its stringent assumptions. We hope that relaxing those assumptions will offer a route to allow these ideas to be applied more broadly.

The first requirement is access to an input distribution, which is often lacking. This can be avoided by using real neural data as the input. Or, if neural data is limited, generative modelling could be used to fit the neural data distribution and new samples drawn from that distribution. Finally, one could imagine a single meta-learner that creates not only the label, but also the data. That is, the meta-learner could generate the entire dataset by transforming a noise sample into an input-output pair. This would have to be carefully regularised to avoid trivial input distributions, but could in principle learn the input statistics that particular networks are tuned to process.

Next, we could relax our second assumption, 
that the learner performs supervised learning. This is a standard assumption in theoretical neuroscience \citep{sorscher2022neural,hiratani2022developmental,schaffer2018odor}, and is often reasonable. Some circuits contain explicit supervision or error signals, like the fly mushroom body or the cerebellum \citep{shadmehr2020population}, and generally brain areas that make predictions (i.e., all internal models), can use their prediction errors as a learning signal. Alternatively, some circuits are well modelled as one area providing a supervisory signal for another, as in classic systems consolidation~\citep{mcclelland1995there}, or receiving supervision from a past version of themselves through replay~\citep{van2020brain}. Nevertheless, much biological learning seems to be unsupervised. Our framework could be extended to these settings by assuming an unsupervised objective and meta-learning the dataset on which an observed circuit performs well. For example, the unsupervised objective might be to produce a lower-dimensional representation of the data with the same dot-product similarity structure as the inputs. If the learner was doing PCA, then the meta-learner would learn data that lay in a linear subspace. It would be interesting to try this for different dimensionality reduction algorithms, or to see which bits of input structure biological unsupervised networks are tuned to process. 

Despite our optimism for this approach, there remain challenges. Most fundamentally, arbitrary functions are still hard to interpret. Our tool turns one problem - what is the inductive bias of this learner - into another - can I intepret this function the learner finds easy-to-generalise?  The second problem is hard but tractable, and is an active area of research \citep{montavon2018methods}; and we hope progress in this area will help our tool. In the meantime, we have shown how our tool provides insight in a variety of settings: for low-dimensional inputs (Fig.~2 - 5), by comparing to ground truth labels (Fig.~6), or by projecting the learnt functions onto an appropriate basis (Fig.~7).

To conclude, the inductive bias is a promising angle from which to understand learning algorithms. Analytic bridges between circuit design and inductive bias have already `explained' the presence of aspects of the circuit through their effect on the network's generalisation properties in both artificial \citep{canatar2021spectral, bahri2021explaining} and biological \citep{bordelon2021population, pandey2021structured, harris2019additive, xie2022task} networks. However, these techniques require very constraining assumptions. We have dramatically loosened these assumptions and shown our tools utility in, among other things, interpreting connectomic data. We believe it will prove useful on other datasets and problems.



\subsection*{Acknowledgements}

The authors thank Cengiz Pehlevan for the conversations that inspired this work; Basile Confavreux \& Blake Bordelon for insightful discussions; Robert Tjarko Lange for very helpful conversations about black-box optimsation; Rodrigo Carrasco-Davis for patient listening and commenting; Clémentine Dominé for commenting on this manuscript; and Pierre Glaser for his exemplary sorting of stressed late-night errors; and Peter Vincent for his help.

We thank the following funding sources: Gatsby Charitable Foundation to W.D. \& P. E. L.; Wellcome Trust (110114/Z/15/Z) to P.E.L.; and Simons Collaboration on the Global Brain Undergraduate Research Fellowship to M. Y..

\bibliography{bibliography}
\bibliographystyle{icml2023}
\newpage
\appendix

\section{Meta-Learning a Simple Backprop Trained ReLU Network}\label{app:ReLU}

\begin{figure}[h]
\begin{center}
\includegraphics[width=0.48\textwidth]{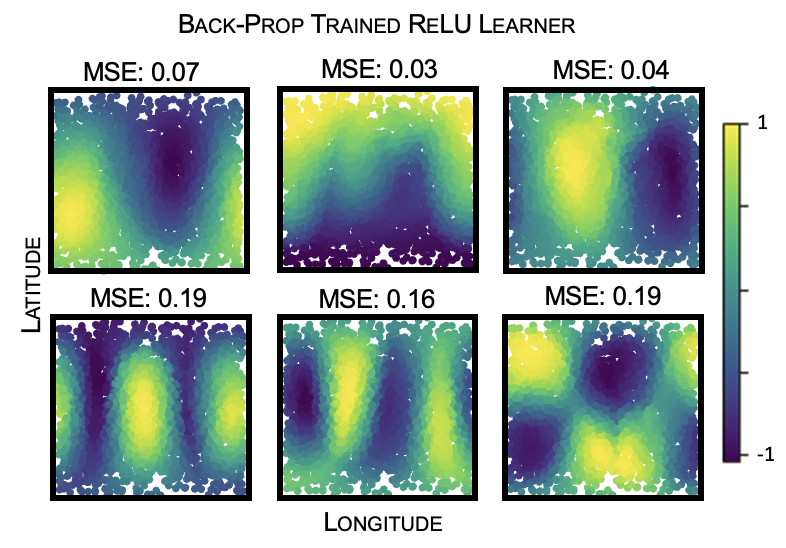}
\caption{\textbf{Meta-learning a Backprop Trained Network} We meta-learn the functions a 2-layer ReLU network trained using backprop finds easy to generalise, as can be seen, this shows a low-frequency bias.}
\label{Fig: 5}
\end{center}
\vspace{-1.5em}
\end{figure}

A simple test of our framework is a feedfoward ReLU network with 2 hidden layers, learnt using gradient descent. While the functions this network finds easy to generalise cannot be extracted analytically, our tool finds that, unsurprisingly, these networks are biased towards smooth explanations of the data, learning six smooth orthogonal classifications that increase in frequency and mean squared error Fig.~\ref{Fig: 5}B. We include this example as the simplest PyTorch implementation of learning the inductive bias of a network trained by gradient descent, in the hope that the code can be easily adapted for future use.

\section{Using Different Divergence Measures}

\begin{figure}[h]
\begin{center}
\includegraphics[width=0.5\textwidth]{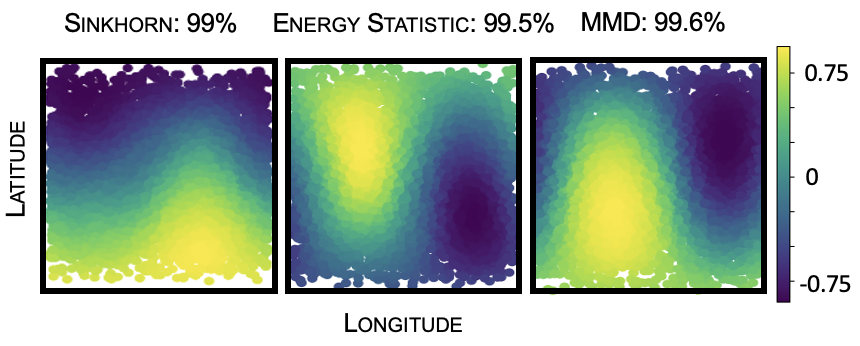}
\caption{We use three different divergence metrics to penalise the meta-learner's label distribution for deviating from uniform between -1 and 1. We add these to the meta-loss, along with the generalisation error of the simple kernel learner introduced in section 2. For all three divergence metrics the meta-learner learns a very close approximation to a first order spherical harmonic, as predicted.}
\label{Fig:Div}
\end{center}
\end{figure}

To persuade the meta-learner to find non-trivial functions, we include a divergence loss that forces the meta-learner's label distribution to take a particular form: uniform between -1 and 1. In this section we show that the particular divergence that we use has little impact on the solutions we find for learning the easiest-to-generalise function of the kernel learner, figure~\ref{Fig: 2}C. Figure~\ref{Fig:Div} shows that a variety of divergence metrics can be used, sinkhorn, an energy statistic from~\citet{szekely2013energy}, and the maximum mean discrepancy from \citet{gretton2012kernel} (implementations from~\citet{josipd2020}). In each case the learnt function has over 99\% norm in the space of first order spherical harmonics, demonstrating that the meta-learner has learnt appropriately.

Other techniques also work well, including adding terms to the loss that force the variance of the meta-learner's label distribution to be 1, which is what we used in figure 8.

\section{Random Re-Seeding cannot Replace Orthogonalisation}

We introduced the orthogonalisation procedure in section~\ref{sec:orthog} so that successive meta-learners have to explore larger areas of function space. It is legitimate to wonder whether simply re-running the optimisation would have had the same effect. Here we show it does not, and orthogonalisation is needed to explore the learner's inductive bias fully.

\begin{figure}[h]
\begin{center}
\includegraphics[width=0.5\textwidth]{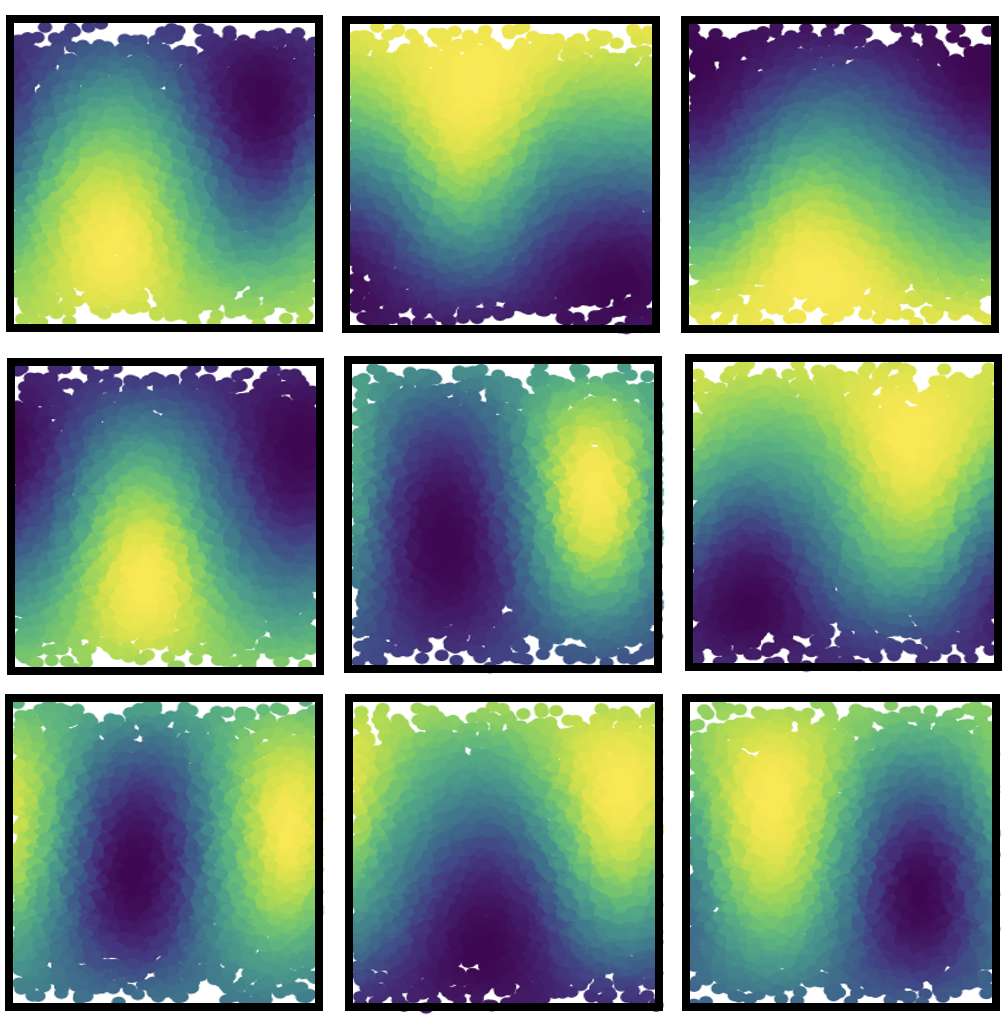}
\caption{Re-running the meta-learner 9 times on the kernel regression algorithm from section 2 produces approximations to the first order spherical harmonics, but no second order functions.}
\label{Fig:Repeat}
\end{center}
\end{figure}

We re-run the meta-learner optimisation without any orthogonality terms for the kernel learner described in section 2. In figure~\ref{Fig:Repeat} we find that the meta-learners find different functions, but only approximations to the first order spherical harmonics. This makes sense, the meta-learner is tasked with finding the easiest-to-generalise non-constant function. For this particular learner there is a degenerate space of such functions and so re-running the meta-learner simply draws another sample from this space of functions. However, to access the second order spherical harmonics that this learner still learns, just worse, we need something like the orthogonality constraint.

\section{Impact of Meta-Learner Architecture on Extracted Functions}

We chose the meta-learner's architecture to be a slightly larger version of the learner's. Our motivation for this is that we want the function class of the meta-learner to be a super-set of the learner's, so that it can learn all the functions the learner could plausibly generalise well.

\begin{figure*}[h]
\begin{center}
\includegraphics[width=\textwidth]{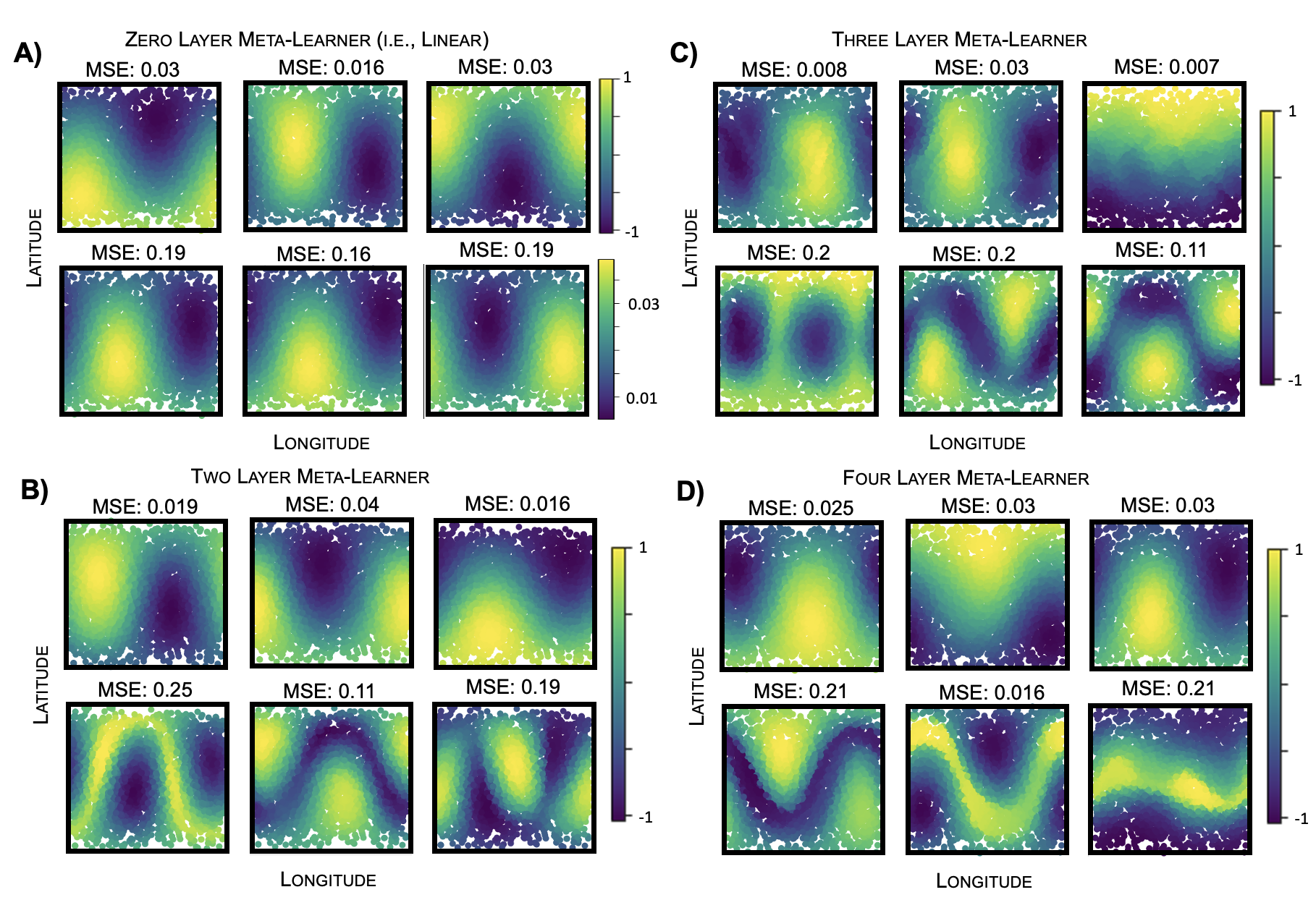}
\caption{We meta-learn 6 easy-to-generalise functions for a simple ReLU network using meta-learners of different depths. This process fails to find more than 3 orthogonal functions when the meta-learner is linear {\textbf (A.)} (note the small label spread, a symptom of a failure to learn), but meta-learners with {\textbf (B.)} 2, {\textbf (C.)} 3, or {\textbf (D.)} all find qualitatively similar results}
\label{Fig:MetaArch}
\end{center}
\end{figure*}

We tested how robust our results were to architectural changes in the meta-learner. We used the simple feedforward 2-hidden layer ReLU network as in Appendix A, trained by backpropagation, and learnt 6 orthogonal functions that the learner finds easy to generalise. We took our meta-learner to be a similar feedforward network of different depth from 4 to 0 hidden layers. Figure~\ref{Fig:MetaArch} shows that the specific choice of meta-learner didn't matter for meta-learners with between 2 to 4 layers. Each learnt 3 approximations to first order spherical harmonics, and 3 to second order harmonics. But a linear learner can't learn more than three orthogonal functions, so failed to find more than three orthogonal generalisable functions.

\section{Connectomic Details}

The networks studied in section 5.3 all have 5 input neurons, a large hidden layer of 150 neurons, and a readout neuron. We train the networks end-to-end to minimise the mean square error with one key architectural difference. Each network is constrained to have a sparse input-to-hidden connectivity: each hidden neuron is connected to only two input neurons, these weights are allowed to vary, all other input-to-hidden weights remain fixed at 0. Figure 7 shows results from three networks that all differ only in how each hidden neurons' connections to the input neurons are chosen. In the simplest network (Random 2-Sparse, Figure 7B) these are chosen randomly; each hidden neuron is equally likely to connect to any pair of inputs. For the second model (Figure 7C) we introduce a bias in the connectivity, one input is much more likely to be chosen than the others, the samples are drawn without replacement from a categorical distrubtion with values 0.6, 0.1, 0.1, 0.1, 0.1. Finally for the last model we introduce grouped connectivity: the inputs come in two groups, if a hidden neuron is connected to one member of a group it is likely to be connected to another. We implement this by dividing the five input neurons into one group of two, and another of three. Each hidden neuron's first connection is drawn randomly from the five. Then the second connection is chosen differently depending on the group of the first: if it connected to a member of the first group with two embers the second connection is chosen with probabilities 0.85, 0.05, 0.05, 0.05; else it is chosen with probabilities 0.1, 0.1, 0.4, 0.4.

\end{document}